\documentstyle[aps,prb,epsf]{revtex}

\begin{document}
\draft

\title
{
\LARGE \bf
 Quantum Depinning of a Pancake-Vortex
from a Columnar Defect}
\author{{\bf D. A. Gorokhov} {\bf and}  {\bf G. Blatter}\\
{\it Theoretische Physik, ETH-H\"onggerberg,
CH-8093 Z\"urich, Switzerland}}
\maketitle
\begin{abstract}
We consider the problem of the depinning of a 
weakly driven ($F\ll F_{c}$) pancake
vortex from a columnar defect
in a Josephson-coupled superconductor,
where $F$ denotes the force acting on the vortex 
($F_{c}$ is the critical force).
 The dynamics of the vortex is supposed to be of the Hall type.
The Euclidean action $S_{\rm Eucl}(T)$ is calculated in the entire
temperature range; the result is universal and
does not depend on the detailed form of the pinning potential.
We show that the transition from quantum to classical
behavior is second-order like with 
the temperature $T_{c}$ of the transition 
scaling like
$F^{{4}/{3}}.$ Special attention is paid to the regime of 
applicability of our results, in particular, the 
influence of the large vortex mass 
appearing in the superclean limit is discussed. 
  \end{abstract}

\hskip1.5cm{preprint ETH-TH-97/22; accepted for publication in Phys. Rev. B}

\hskip1.5cm e-mail: gorokhov@itp.phys.ethz.ch

\section{Introduction}

In recent years, quantum creep
 of vortices in high $T_{c}$-superconductors has attracted considerable
interest 
as, on the one hand, this phenomenon is responsible
for the dissipation of energy and thus is relevant
from a technological point of view, while, on the other hand,
it represents an interesting example of a macroscopic quantum phenomenon,
the theoretical study of which is challenging.
A particularly well defined and technologically relevant configuration
are vortices trapped by columnar defects introduced into the sample
by heavy ion irradiation. Measurements of the critical current
density and the magnetic relaxation rate show the strong
influence these pinning centers have on the vortex dynamics\cite{experim}.
In this paper we consider the problem of the 
depinning of a pancake vortex governed by Hall dynamics 
from a columnar defect in a layered superconductor
in the presence of a small ($j\ll j_{c}$) transport current.
The external magnetic field is chosen parallel
to the $c$-axis of the superconductor.
In the limit $j\rightarrow 0$ the problem is 
semiclassical and the Euclidean action can be calculated in the 
whole temperature range. 
In the limit $T=0$ the problem discussed 
above has been considered in Ref.\cite{Bulaevskii},  
however, 
it appears that the approximations made are too rough, leading to 
an inexact result
for the decay rate. The main goal of the present work is to 
improve on the analysis
of Ref.\cite{Bulaevskii} and to extend the calculations to
the entire temperature range.
We adopt the semiclassical approach instead of the lowest Landau level
(LLL) approximation used in Ref.\cite{Bulaevskii}.
For a general review concerning the decay of metastable states
see 
the review of H\"anggi, Talkner, and Borkovec, see 
Ref.\cite{Haenggi}. 
The problem of quantum and classical Hall creep of vortices in various
geometries and for different driving forces $F\alt F_{c}$ and
$F\ll F_{c}$ has been studied by various authors, see 
Refs\cite{Feigel'man,Ao,Stephen,Chudnovsky3,Sonin,Morais-Smith,Gorokhov2}.

The process of quantum tunneling 
is described by a time-dependent 
saddle-point solution. 
Consequently, the calculations of the
decay rate require the specification of the vortex dynamics.
The 
different contributions to the 
dynamics considered traditionally are of the 
massive, the dissipative, and the Hall type.
The (low frequency) equation of motion of a  single vortex can be written 
in the form
\begin{equation}
\frac{\Phi_{0}}{c}{\bf j}\wedge{\bf n}
-{\bf \nabla}U_{pin}
=-\eta {\bf v}+m{\bf \dot v}+
\alpha{\bf v}\wedge {\bf n}, 
\label{dwizhenie}
\end{equation}
where the dynamical forces are balanced by the Lorentz and pinning
forces.
In conventional superconductors
one can neglect the contribution of the mass and Hall terms,
whereas in high-$T_{c}$ superconductors the Hall force may
 become relevant.  
In particular,
it is widely believed\cite{Blatter} that at low temperatures
the superclean limit can be reached
where the Hall term is large.
In this case $\omega_{0}\tau\agt 1,$ where $\omega_{0}$ is the level
spacing inside the vortex core and $\tau$ is the quasiparticle relaxation
time.
Indeed, recent
Hall angle measurements\cite{Matsuda,Harris}  
demonstrate that
the limit $\alpha\agt\eta$ can be realized. 
On the other hand, the large parameter $\omega_{0}\tau$
also gives rise to a large vortex mass. Microscopic calculations
show that the vortex mass is enhanced
by a factor ${\left (\epsilon_{F}/\Delta\right )}^{2}\sim 100$
in comparison to the dirty limit\cite{Kopnin,Simanek,AVO}.
Still, it can be shown that for frequencies $\omega < \omega_{0}$
the Hall force wins over the inertial one, whereas at high temperatures
$\omega > \omega_{0}$ the vortex equation of motion 
cannot be cast into the simple form (\ref{dwizhenie})
(an accurate description produces dispersive transport coefficients
$\eta (\omega ),$ and $\alpha (\omega )$). 
In this situation a good
starting point is to ignore the contribution of the vortex mass and solve
the remaining Hall tunneling problem. 
The advantage of this treatment
lies in
 the fact that in the limit $j\ll j_{c}$ the problem allows for an
analytical solution in the whole temperature range with a universal
answer: 
the Euclidean action depends only on the depth of the pinning potential,
the detailed shape of the potential being irrelevant.
In a second step we establish the consistency of this approximation
in the physically relevant regime of parameters.
The outline of the paper is as follows: In section \ref{model}
we discuss the model and the qualitative picture of the tunneling
process.
In section 
\ref{problem} we calculate the Euclidean action in the whole temperature range
and show that the problem always exhibits a second-order like transition
from quantum to classical behaviour. 
Furthermore, we provide estimates for the preexponential factors in the
various regimes.
Finally, in  section
\ref{conclusion} we discuss the conditions of applicability of
our results.

\section{Model and qualitative picture}
\label{model}

Consider a pancake vortex 
with a Hall-force dominated 
dynamics trapped in a 2D
potential well 
$U_{0}\left (\sqrt{x^{2}+y^{2}}\right )$ 
and subject to an external
force $F,$ i.e., the effective potential $U\left (x,y\right )$
takes the form
\begin{equation}
U\left (x,y\right)=
U_{0}\left (\sqrt{x^{2}+y^{2}}\right )-Fx.
\label{potential}
\end{equation}
The function $U_{0}\left (r\right )$ is supposed to be monotonously
increasing, $U_{0}\left (0\right )=0$ and
$U_{0}\left (\infty \right )=U_{0}.$ At distances $r$ much larger than
the characteristic radius  $a$
of the pinning potential 
but still smaller than the magnetic field penetration length $\lambda_{ab}$
in the $ab$-plane,
$U_{0}\left (r \right )$ behaves as $U_{0}-{B}/{r^{2}}$ with
$B$ a constant of order $U_{0}a^{2}$ (see Ref.\cite{Mkrtchyan}).

\vskip0.5cm
\centerline{\epsfxsize=14cm \epsfbox{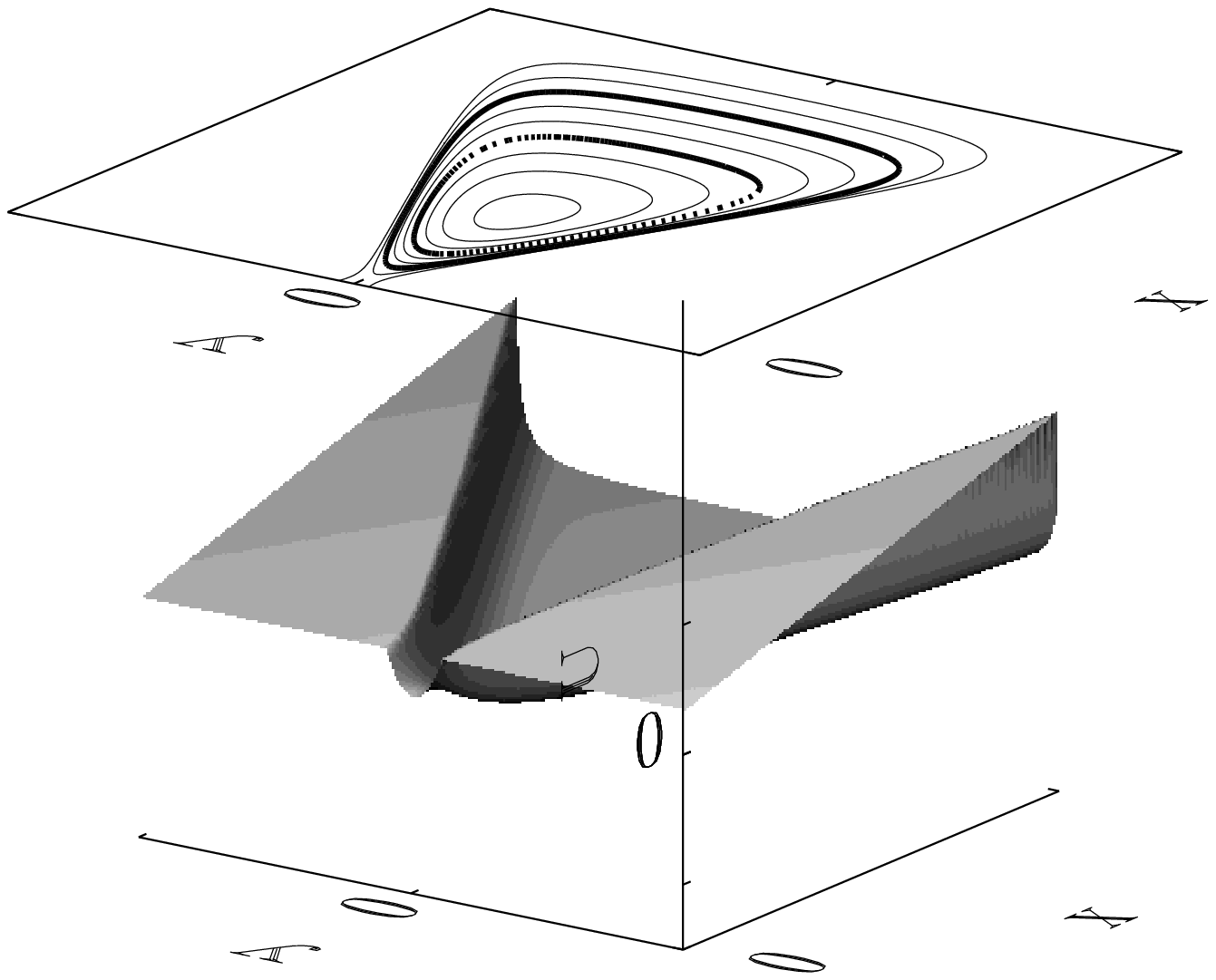}}
{\footnotesize {\bf Fig.1}~
Inverted potential $U$ after the transformation $t\rightarrow -i\tau,$
$y\rightarrow iy.$
Note that the inversion affects only the unstable direction
along the $x$-axis. The equipotential lines define the quasiclassical
trajectories. 
The thick solid line corresponds to the zero
temperature instanton,
 whereas the thick dotted line  is a finite temperature
bounce trajectory.}
\vskip0.3cm
\hskip-0.7cm

The present problem is equivalent to that of the 
motion of a charged particle in a strong magnetic field.
In this case, 
the appropriate Lagrangian takes the form
\begin{equation}
L=\alpha {\dot x} y -U(x,y).
\label{Lagrangian}
\end{equation}
The 
precise
conditions allowing us to neglect the mass of the particle
will be discussed in section \ref{conclusion}.
The particle whose motion is described by the Lagrangian~(\ref{Lagrangian})
moves along the equipotential lines of the potential 
landscape $U(x,y).$
The Euclidean action of the particle 
as obtained through the substitution 
$S=\int Ldt\rightarrow -iS$ and
$t\rightarrow -i\tau$
can be written in the form
\begin{equation}
S_{\rm Eucl}=
\int_{-{\hbar}/{2T}}^{+{\hbar}/{2T}}
\left [-i\alpha{\dot x}{y}+U(x,y)\right ]d\tau.
\label{imaginary}
\end{equation}
Obviously, the imaginary unit appears in the Euclidean action,
i.e., the saddle-point solution is in general complex.
However,  in the case studied here it is possible to reduce the complex
problem to a real one\cite{Feigel'man,Blatter} via performing the additional
transformation $y\rightarrow iy.$ If the potential
$U(x,y)$ satisfies the condition ${\rm Im}\left\{U(x,iy)\right \}=0,$ 
we obtain a real-time problem. In addition, if 
after the $y\rightarrow iy$-transformation the Lagrangian~(\ref{imaginary})
exhibits a saddle-point solution, we can find it and calculate the
decay rate. One can easily see from Eq.~(\ref{potential})
that the condition ${\rm Im}\left\{U(x,iy)\right \}=0$ is 
indeed
satisfied for our
potential,\cite{note1} 
i.e., we can study the effective problem of the tunneling of a particle
whose dynamics is described by the Euclidean action
\begin{equation}
S_{\rm Eucl}[x(\tau ),y(\tau )]=\int_{-{\hbar}/{2T}}^{+{\hbar}/{2T}}
\left [
\alpha{\dot x}y+U_{0}\right (\sqrt{x^{2}-y^{2}}\left )-Fx\right ]d\tau.
\end{equation}
In Fig.~1 we show the inverted potential after the transformations
$t\rightarrow -i\tau$ and $y\rightarrow iy$ together with the equipotential
lines. Obviously, this construction produces a new potential
shape exhibiting a bounce solution (saddle-point). Note that the inversion
affects the potential only in the unstable direction
along the force.

As the potential energy $U(x,iy)$ 
 is preserved during
motion, the imaginary time trajectories satisfy the equation
$U_{0}\left (\sqrt{x^{2}-y^{2}}\right )-Fx={\rm Const}.$
In Fig.~2
we plot these trajectories for the problem at hand.
At zero temperature and in the limit $F\rightarrow 0$ we obtain
$U_{0}\left (\sqrt{x^{2}-y^{2}}\right )-Fx=0,$
i.e., 
$S_{\rm Eucl}=\alpha \oint ydx=\alpha A,$
where $A$ is the area encircled during the periodic motion.
One can easily see from Fig.~2 that in the limit 
$F\rightarrow 0$ the encircled area is equal to that of a triangle
$CA A^{\prime}.$ With $CB\simeq{U_{0}}/{F}, A A^{\prime}\simeq{2U_{0}}/{F},$ 
we obtain  the area 
$A={\left ({U_{0}}/{F}\right )}^{2}$ and the Euclidean action
is given by the expression 
$S\left(0\right )=\alpha{\left ({U_{0}}/{F}\right )}^{2}$
(in Ref.\cite{Bulaevskii} the result $S\left (0\right )=
\left ({23}/{5}\right )\alpha{\left ({U_{0}}/{F}\right )}^{2}$ has been obtained, in contradiction with the above analysis).
In the next section we shall generalize the zero temperature result
to the case of arbitrary temperatures.

\centerline{\epsfxsize=7cm \epsfbox{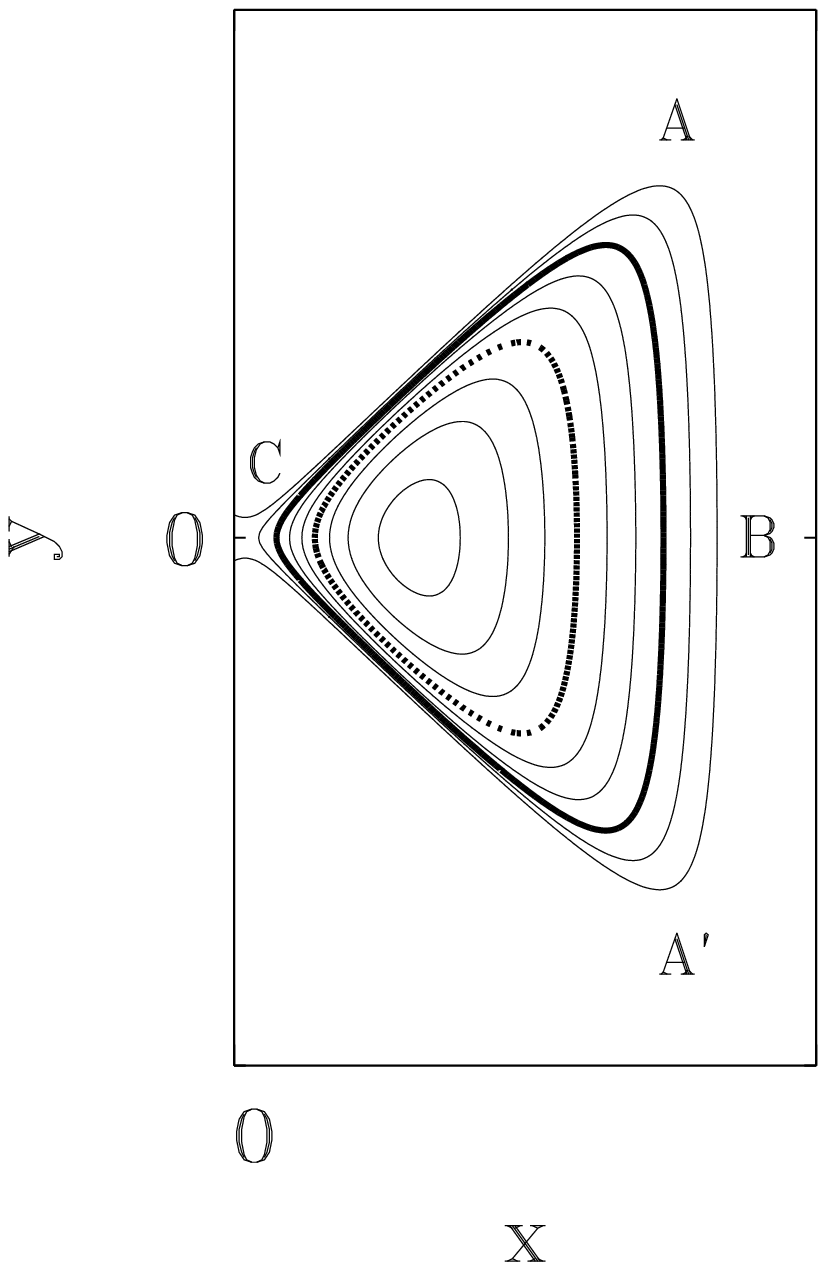}}
{\vskip0.5cm\footnotesize {\bf Fig.2}~Quasiclassical $(x,y)$-trajectories corresponding to 
the tunneling of the vortex. The Euclidean action corresponding
to the zero temperature trajectory 
(thick line)
is equal to the encircled area.
 The thick dotted line  marks a finite temperature
bounce trajectory..}
\vskip0.3cm
\hskip-0.7cm

\section{Decay rate}
\label{problem}
\subsection{Euclidean action}

As has been shown by Volovik\cite{Volovik} 
(see also Refs.\cite{Feigel'man,Blatter}), 
the motion of a massless particle
in a magnetic field subject to a potential $U\left (x,y\right )$
is equivalent to the 
1D dynamics
 of a particle described by the Hamiltonian
$U\left (x,{p}/{\alpha}\right ).$
Consequently, our original 2D problem (\ref{Lagrangian}) 
reduces to the 1D problem
with the Hamiltonian given by the expression
\begin{equation}
H\left (x,p\right )=
U_{0}\left (\sqrt{x^{2}+{\left ({p}/{\alpha}\right )}^{2}}\right )-Fx.
\label{Hamiltonian}
\end{equation}
Let us show that in the limit $F\rightarrow 0$
the problem is semiclassical.
The semiclassical wave function can be written in the form\cite{Landau}
\begin{equation}
\Psi \left (x \right )=
\frac{C}{\sqrt{\dot x}}
\exp{\left (\frac{i}{\hbar}\int\limits_{x^{\prime}}^{x}p\thinspace dx \right )}=
C^{\prime}\exp{\left (\int\limits_{x^{\prime}}^{x} dx
\left (\frac{ip}{\hbar}-\frac{1}{2}
\frac{{{\partial}^{2}H}/{\partial p^{2}}}{{\partial H}/{\partial p}}
\frac{dp}{dx} \right )\right )}.
\end{equation}
with $x^{\prime}$ an integration constant.
The semiclassical approximation is applicable if
\begin{equation}
\left |p\right |\gg\hbar
\left |
\frac{{{\partial}^{2}H}/{\partial p^{2}}}{{\partial H}/{\partial p}}
\frac{dp}{dx}\right |.
\end{equation}
Using Eq.~(\ref{Hamiltonian})
we can write for
${\partial^{2}_{p}H}/{\partial_{p}H}$
\begin{equation}
\left |
\frac{{{\partial}^{2}H}/{\partial p^{2}}}{{\partial H}/{\partial p}}
\right |
=
\frac{1}{\left |p\right |}
\left |
\frac{x^{2}}{x^{2}+{\left ({p}/{\alpha}\right )}^{2}}+
\frac{{\left ({p}/{\alpha}\right )}^{2}}
{\sqrt{x^{2}+{\left ({p}/{\alpha}\right )}^{2}}}
\frac{{{\partial}^{2}U_{0}}/{\partial u^{2}}}{{\partial U_{0}}/{\partial u}}
\right |,
\end{equation}
where $u={\left ({x^{2}+{\left ({p}/{\alpha}\right )}^{2}}\right )}^{{1}/{2}}.$
If $\left |x^{2}+{\left ({p}/{\alpha}\right )}^{2}\right |\ll a^{2},$
we can use $U_{0}\cong{{\kappa}u^{2}}/{2},$
i.e., 
$
\left |
{\partial^{2}_{p}H}/{\partial_{p}H}
\right |
\cong {1}/{\left |p\right |}$ in the above limit.
If $
\left |
x^{2}+{\left ({p}/{\alpha}\right )}^{2}
\right |\gg a^{2},$
\begin{equation}
{\partial^{2}_{p}U_{0}}/{\partial_{p}U_{0}}
\sim{\left ({{x^{2}+{\left ({p}/{\alpha}\right )}^{2}}}
\right )}^{-{1}/{2}},
\end{equation}
and we obtain 
$
\left |
{\partial^{2}_{p}H}/{\partial_{p}H}
\right |
<{A}/{\left |p\right |},$ 
with $A$ a constant of order one.
Finally, at
$
\left |
x^{2}+{\left ({p}/{\alpha}\right )}^{2}\right |
\sim a^{2},$
we obtain again 
$\left |
{\partial^{2}_{p}H}/{\partial_{p} H}\right |\simeq
{A^{\prime}}/{|p|},$
$A^{\prime}\sim 1.$
Consequently, a sufficient criterion for the applicability of the
 semiclassical approximation takes the form
\begin{equation}
\left |\frac{\hbar}{p^{2}}\frac{dp}{dx}\right |\ll
1,
\label{criterion}
\end{equation}
which is the same criterion as for a ``usual'' Hamiltonian of the form
$H\left (x,p\right )={p^{2}}/{2m}+U\left (x\right ).$

 Using the standard technique
of binding semiclassical wave-functions we obtain the following expression
for the imaginary parts of the metastable energy levels\cite {Galitsky},
\begin{equation}
\Gamma_{n}=
\frac{\omega\left (E_{n}\right )}{4\pi}
\exp{\left (-\frac{2}{\hbar}\int\limits_{c_{n}}^{b_{n}}|p|
\thinspace dx \right )}=
\frac{\omega\left (E_{n}\right )}{4\pi}
\exp{\left (-\frac{S_{n}}{\hbar}\right )},
\label{decay}
\end{equation}
with $E_{n}$ the energy levels at zero driving force,
$\omega (E_{n})$ are the oscillation frequencies, and
$c_{n}$ and $b_{n}$ denote the turning points. 
The decay rate $\Gamma$ can be found by averaging over the Boltzmann
distribution (a general discussion concerning finite temperature
decay and the role of dissipation is found in Ref.\cite{Haenggi}) 
\begin{equation}
\Gamma=\left ({2}/{Z}\right )\sum_{n}\Gamma_{n}e^{-{E_{n}}/{T}},
\label{average}
\end{equation}
with $Z$ the partition function for the case $F=0.$
As the semiclassical approximation is applicable, we can 
substitute the sum in Eq.~(\ref{average}) by an integral
and 
make use of the method of
steepest descent. The extremal equation 
then takes the form ${\partial S}/{\partial E}=-\tau (E),$
with $\tau (E)$ the imaginary time oscillation period. 
Consequently, if one can calculate the function
$\tau (E)$ from the solution of the classical equation of motion,
the function $S (E)$ can be reconstructed via simple integration,
$S(E)=-\int\limits^{E}
\tau (E^{\prime} )d E^{\prime}.$ Let us carry out this program 
for the present problem.

The semiclassical trajectories can be found
as the solution of the equation $H\left (x,p\right )=E,$
with $E$ the energy, i.e.,
\begin{equation}
p (x)=\pm\alpha
\sqrt{f^{2}\left (E+Fx\right )-x^{2}},
\label{momentum}
\end{equation}
with $f=U_{0}^{-1}$ the inverse function of the potential shape
$U_{0}.$
There is a region $c\le x\le b $ ($b>c>0$), where the function
$p(x)$ is purely imaginary
(the $x$-coordinates $c$ and $b$ are associated with
the turning points $C$ and $B$ in Fig.~2).
 The equation 
$f^{2}\left (E+Fx \right )-x^{2}=0$
has two solutions: At small $x,$ $f^{2}\left (E+Fx \right )\cong
\left ({2}/{\kappa}\right )\left (E+Fx\right ),$ i.e.,
$c={F}/{\kappa}+\sqrt{{F^{2}}/{\kappa^{2}}+{2 E}/{\kappa}}.$
As $x\rightarrow {\left (U_{0}-E\right )}/{F},$
$f^{2}\left (E+Fx\right )\rightarrow\infty,$ 
the equation  
$f^{2}\left (E+Fx \right )-x^{2}=0$
has another root $b\cong{\left (U_{0}- E\right )}/{F}$
(the expressions for $c$ and $b$ are applicable for any energy
not too close to $U_{0}$).
 The decay rate of a metastable state
with an energy $E$ is proportional to 
$\exp{\left [-\left ({2}/{\hbar}\right )
\int_{c\left (E\right )}^{b\left (E\right )}
|p| dx \right ]}.$ Note that everywhere inside the interval
$c\le x\le b,$ except for the vicinity of the points $c$ and $b,$
$|p|\cong \alpha x$
and the condition~(\ref{criterion})
is fulfilled. At the points $b$ and $c,$ $p=0.$
These points play the role of  turning points in the ``usual'' semiclassical
approximation\cite{Landau}.
  Consequently, we have shown that in the limit $F\rightarrow 0$
the semiclassical approximation is applicable.

Let us calculate the Euclidean action.
Using Eqs.~(\ref{decay}) and (\ref{momentum})
we can write for $S$
\begin{equation}
S (E)=
2\alpha\int\limits_{c}^{b}
\sqrt{x^{2}-
f^{2}\left (E+Fx\right )}
\thinspace dx.
\end{equation}
The oscillation time $\tau(E)$ satisfies the equation
$\tau (E) = -{\partial S}/{\partial E},$ i.e.,
\begin{equation}
\tau \left (E\right )=
\frac{2\alpha}{F}
\int\limits_{c}^{b}
\frac{x\thinspace dx}
{\sqrt{x^{2}-
f^{2}\left (E+Fx\right )}}
\label{pperiod}
\end{equation}
In the limit $F\rightarrow 0$ 
almost everywhere inside the interval $[c,b]$ we have  
$x^{2}\gg f^{2}\left (E+Fx\right )$
and one can write for the period $\tau (E)$
\begin{equation}
\tau (E)=
\frac{2\alpha}{F}\left (b-c \right )+
C \left (c\right ) 
+
C \left (b\right ), 
\label{period}
\end{equation}
with $C(c)$ 
and $C(b)$ the contributions of the turning points $c$ and $b$
where the function 
$x^{2}-f^{2}\left (E+Fx\right )$ vanishes.
It can be shown that $C(c)$ is relevant only if $E$
is close to $-{F^{2}}/{2\kappa}$ (see below)
 and the contribution of the point $b$ is
always negligible.
Calculating $C(c),$ substituting the result into Eq.~(\ref{period}),
and taking into account that for $E$ not very close to
$U_{0},$ $b-c\simeq {\left (U_{0}-E\right )}/{F},$
we obtain
\begin{equation}  
\tau (E)\simeq\frac{2\alpha}{F^{2}}\left (U_{0}-E\right )+
\frac{2\alpha}{\kappa}
\ln
\frac{2{\tilde x}}{\sqrt{\frac{F^{2}}{\kappa^{2}}+\frac{2E}{\kappa}}},
\label{period1}
\end{equation}
where ${\tilde x}$ is an $E$-independent 
cutoff parameter arising from the integration
in the vicinity of the point $c.$

Next, we need to solve the equation
$\tau (E)={\hbar}/{T}$ and find the energy $E$ of the saddle-point 
trajectory at finite temperature $T.$
The second term in Eq.~(\ref{period1}) is relevant
only if the solution of the equation $\tau (E)={\hbar}/{T}$
is very close to $-{F^{2}}/{2\kappa},$
i.e., at low temperature where the equation
${2\alpha \left (U_{0}-E\right )}/{F^{2}}={\hbar}/{T}$
has no solution
(this is the case when $T<{\hbar F^{2}}/{2\alpha U_{0}}$),
and produces merely an exponentially small correction to the zero temperature
result.
This 
behavior is typical for a Hamiltonian problem where the finite temperature
boundary conditions have a vanishingly small effect on the bounce
solution at small temperatures\cite{Grabert}.
On the other hand, if $T>{\hbar F^{2}}/{2\alpha U_{0}},$
the solution of Eq.~(\ref{period1}) is given by 
$E=U_{0}-{\hbar F^{2}}/{2\alpha U_{0}}$ 
(up to exponentially small corrections).
After a simple integration
$S(E)=-\int\limits^{E}\tau (E^{\prime})dE^{\prime}$
we obtain the Euclidean action
$S_{\rm Eucl}={\alpha{\left (U_{0}-E \right )}^{2}}/{F^{2}}+{E}/{T}$
(the integration
 constant is obtained  from the condition $S(0)=
\alpha {{U_{0}}^{2}}/{F^{2}}$).
In summary,
for temperatures $T<{\hbar F^{2}}/{2\alpha U_{0}}$
the Euclidean action is constant up to exponentially
small corrections. At ${T_{1}}={\hbar F^{2}}/{2\alpha U_{0}},$
$S_{\rm Eucl}$ begins to decrease,  
$
S_{\rm Eucl}\left (T\right )=
{\hbar U_{0}}/{T}-{\hbar^{2} F^{2}}/{4\alpha T^{2}}.
$
We then can write the following expression for the Euclidean action
in the whole temperature interval (about the applicability
of this result to the high temperature regime see below)
\begin{equation}
S_{\rm Eucl}=
\left\{ \begin{array}{r@{\quad\quad}l} 
\displaystyle{\alpha{\left (\frac{U_{0}}{F}\right )}^{2},} 
& {T<{\hbar F^{2}}/{2\alpha U_{0}}\equiv T_{1},}\\ \noalign{\vskip 5 pt}
\displaystyle{\frac{\hbar U_{0}}{T}-{\frac{\hbar^{2}F^{2}}{4\alpha T^{2}}}}, & 
{T>T_{1}.}\end{array}\right.
\label{otwet}
\end{equation}

\subsection{Crossover to Classical Behavior}
Next, let us calculate the crossover temperature $T_{c}$
from the thermal-assisted quantum regime to the purely
thermal activation. Below we use the perturbative procedure
which is applicable only for second-order 
transitions from quantum to classical behavior. 
For a first-order transition this approach breaks down.
However, we will show that if the potential $U_{0}(r)$
satisfies the required conditions ($U(r)$
is monotonously increasing and $U(r)\simeq U_{0}-{B}/{r^{2}},\thinspace
r\rightarrow\infty$)
a second-order transition takes place.
The crossover temperature $T_{c}$ is equal to ${\hbar}/{\tau_{0}},$
where $\tau_{0}$ is the imaginary time oscillation period
of the system in the vicinity of the 
time-independent
thermal saddle-point solution.
This solution is given by the equation $x (\tau )=x_{max}$
with $x_{max}$ the point where the function $U(x,0)$
takes its maximal value. Near this point 
$U(x,y)\simeq U_{0}-{B}/{\left ( x^{2}+y^{2}\right )}-Fx.$
For this dependence the function $\tau (E)$ (see Eq.~(\ref{pperiod}))
can be calculated exactly,
\begin{eqnarray}
& \thinspace & 
\tau(E) =\frac{2\alpha B}{F^{2}}
\frac{1}{{\left (b+d \right )}^{{3}/{2}}
{\left (b+c\right )}
{\left (b-d \right )}^{{1}/{2}}}
\Bigg [
(b+d)
F\left (\frac{\pi}{2},
\sqrt{
\frac{(b-c)(b+c)}{(b-d)(b+d)}}
\right )\nonumber\\
& \thinspace & 
+\left (c- d\right )\Pi \left (\frac{\pi}{2},
\frac{(b-c)(b+c)}{(b-d)(b+d)},
\sqrt{
\frac{(b-c)(b+c)}{(b-d)(b+d)}}
\right )
\Bigg ],
\label{crossover}
\end{eqnarray}
where $b\ge c>d$ are the three roots of the equation
\begin{equation} 
U_{0}-\frac{B}{x^{2}}-Fx=E.
\label{as}
\end{equation}
Again $b$ and $c$ are the turning points of the imaginary time
trajectory,
\begin{equation}
\Pi\left ({\pi}/{2},n,k\right )=\int_{0}^{{\pi}/{2}}
{d\phi}
{\left [{\left (1-n\sin^{2}\phi\right )\sqrt{1-k^{2}\sin^{2}\phi}}
\right ]}^{-1}
\end{equation}
 is the complete
elliptic integral of third order and 
\begin{equation}
F\left ({\pi}/{2},k\right )=
\int_{0}^{{\pi}/{2}}
d\phi
{\left [1-k^{2}\cos^{2}\phi\right ]}^{-{1}/{2}}
\end{equation}
 is 
the complete elliptic integral of second order.
Here we consider only the physically relevant 
case $E\le E_{max}=U_{0}-\left ({3}/{2^{{2}/{3}}}
\right  )B^{{1}/{3}}F^{{2}/{3}}.$ If $E=E_{\rm max},$  
$b=c=x_{max}={\left ({2B}/{F}\right )}^{{1}/{3}},$
and $d=-{1}/{2^{{2}/{3}}}{\left ({B}/{F}\right )}^{{1}/{3}}.$ 
Substituting these values into 
Eq.~(\ref{crossover}) we obtain
\begin{equation}
T_{c}=\frac{\sqrt{3}}{2^{{4}/{3}}\pi}
\frac{\hbar F^{{4}/{3}}}{\alpha B^{{1}/{3}}}.
\label{Tc}
\end{equation}
Inserting $T>T_{c}$ into Eq.~(\ref{otwet}) 
in the limit $F\rightarrow 0$
we obtain
$S_{\rm Eucl}\left (T>T_{c}\right )\simeq {U_{0}}/{T},$
such that we 
can use the result (\ref{otwet})
at any temperature in this limit.
The dependence $S_{\rm Eucl}(T)$ is plotted in
Fig.~3. 

Let us show that the problem  
exhibits a second-order like transition
at $T=T_{c}.$
The 1D Hamiltonian system produces a smooth transition at $T_{c}$
if its imaginary time oscillation period $\tau (E)$ is a monotonous function 
of energy\cite{Lifshitz,Chudnovsky,Chudnovsky1,Gorokhov}.
A simple criterion to verify the monotonicity of the function
$\tau (E)$ is given by the derivative $\partial_{E} \tau $
evaluated at $E_{\rm max}$
(see also Ref.\cite{Gorokhov}):
 For $\partial_{E}\tau |_{E_{\rm max}}<0$
the function is monotonous and we have a second-order like transition.
If $b\simeq {\left (U_{0}-E\right )}/{F}\gg c ,$
(i.e., for small $F$ and $E$ not too close to $E_{max}$),
one can use Eq.~(\ref{period}) 
 to show that the function $\tau (E)$ is monotonous in this interval.
On the other hand, one can use Eq.~(\ref{crossover})
as long as the roots $b$ and
$c$ greatly exceed the characteristic radius $a$ of the pinning potential.
In the limit $F\rightarrow 0$
and for $E$ close to $E_{max},$
 $b,c\sim {({B}/{F})}^{{1}/{3}},$
and the condition $b,c\gg a$ is well satisfied.

\vskip0.0cm
\centerline{\epsfxsize=16cm \epsfbox{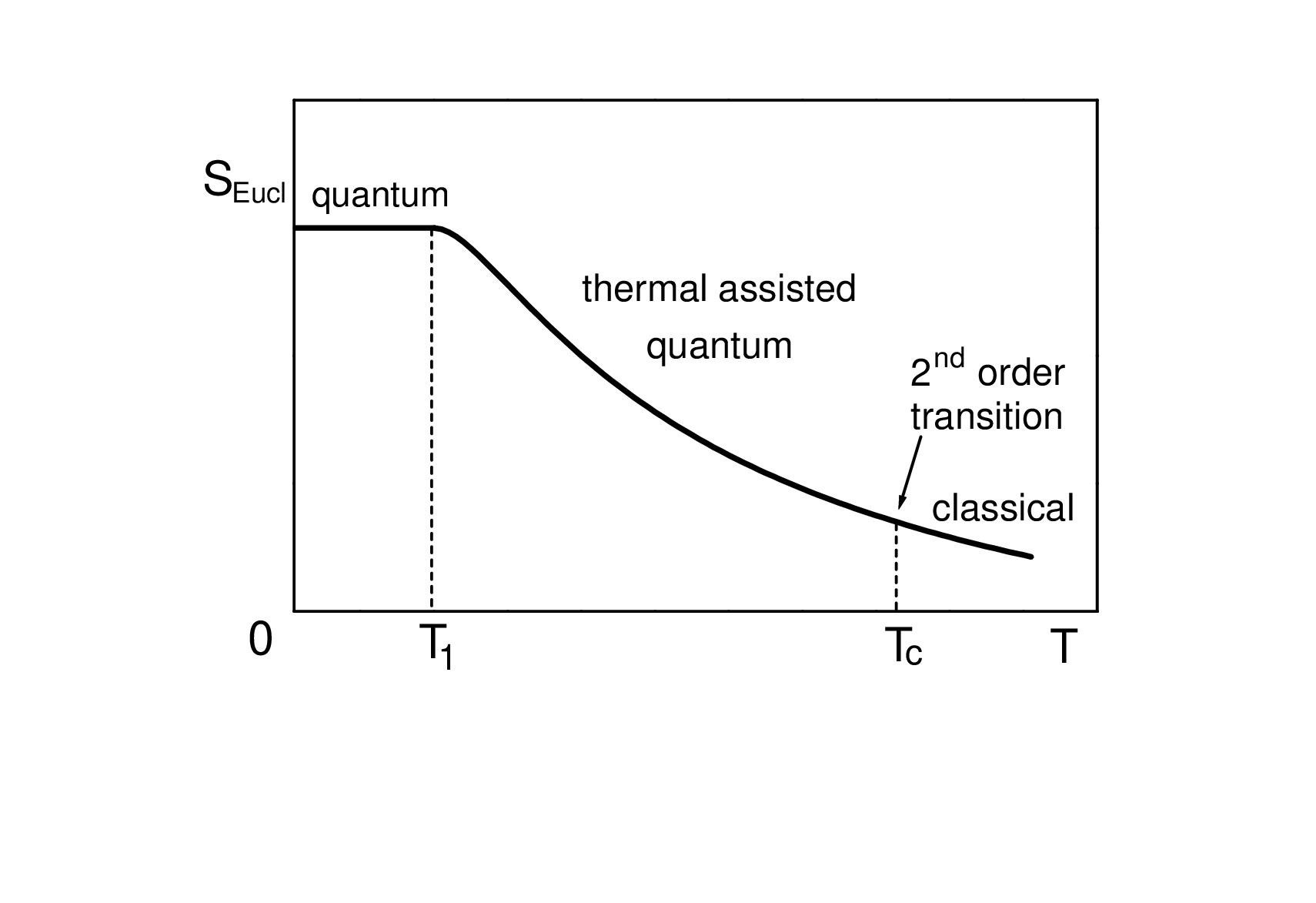}}
\vskip-2.5cm{\footnotesize {\bf Fig.3}~Euclidean 
action as a function of temperature. At $T<{ T_{1}}=
{\hbar F^{2}}/{2\alpha U_{0}},$ $S_{\rm Eucl}$ is a constant up to 
exponentially small corrections. In the regime 
$T>{ T_{1}}$ the Euclidean action
begins to decrease, 
 see Eq.~(\ref{otwet}). The temperature $T=T_{c}$ 
(see Eq.~(\ref{Tc}))
marks the second-order like 
transition from quantum to classical behavior.
In the vicinity of $T_{c}$ Eq.~(\ref{vicinity}) gives an accurate
description of the action.}
\vskip0.3cm
\hskip-0.7cm

We reparametrize the energy in the form 
\begin{equation}
E=U_{0}-\lambda{B}^{{1}/{3}}F^{{2}/{3}}
\label{lambda}
\end{equation}
where $\lambda=\lambda (E)$ is a dimensionless parameter. Substituting  
this expression into Eq.~(\ref{as}) we obtain for large $\lambda$
($E$ away from $E_{max}$),
$c\simeq{\lambda}^{-{1}/{2}}{({B}/{F})}^{{1}/{3}},$ and
$b\simeq\lambda{({B}/{F})}^{{1}/{3}},$ i.e., $b\gg c$
for $\lambda\agt 5$ and Eq.~(\ref{period1}) 
is valid in this region.
Consequently, if we show that the oscillation period
given by Eq.~(\ref{crossover}) is a monotonous function of $\lambda$
for $\lambda<5$,
a second-order transition from quantum to classical behavior takes place.
The function (${\tilde \tau}=
\left ({\tau F}/{2\alpha }\right )
{\left ({F}/{B} \right )}^{{1}/{3}}$)
\begin{eqnarray}
& \thinspace & 
{\tilde \tau}(\lambda ) =
\frac{1}{{\left ({\tilde b}+{\tilde d} \right )}^{{3}/{2}}
{\left ({\tilde b}+{\tilde c}\right )}
{\left ({\tilde b}-{\tilde d} \right )}^{{1}/{2}}}
\Bigg [
({\tilde b}+{\tilde d})
F\left (\frac{\pi}{2},
\sqrt{
\frac{({\tilde b}-{\tilde c})({\tilde b}+{\tilde c})}
{({\tilde b}-{\tilde d})({\tilde b}+{\tilde d})}}
\right )
\nonumber\\
& \thinspace & 
+\left ({\tilde c}- {\tilde d}\right )\Pi \left (\frac{\pi}{2},
\frac{({\tilde b}-{\tilde c})({\tilde b}+{\tilde c})}
{({\tilde b}-{\tilde d})({\tilde b}+{\tilde d})},
\sqrt{
\frac{({\tilde b}-{\tilde c})({\tilde b}+{\tilde c})}{({\tilde b}-{\tilde d})
({\tilde b}+{\tilde d})}}
\right )
\Bigg ],
\label{crossover1}
\end{eqnarray}
is plotted in  Fig.~4 (solid line); 
here ${\tilde b}\ge {\tilde c}>{\tilde d}$ are the roots of
the equation ${1}/{x^{2}}+x=\lambda .$
We can see that the function ${\tilde \tau}(\lambda )$
is monotonously increasing. Surprisingly, even for 
$\lambda =\lambda_{\rm min}
\simeq {3}/{2^{{2}/{3}}},$ the slope of ${\tilde \tau}$ is 
close to its asymptotic value ${{\tilde \tau}(\lambda })\sim \lambda + 
{\rm Const}, 
\thinspace\lambda
\rightarrow\infty .$  
Close to the point $\lambda_{min}={3}/{2^{{2}/{3}}},$ we find
\begin{equation}
{\tilde\tau }(\lambda )\simeq 
\frac{2^{{1}/{3}}}{\sqrt{3}}\pi+\frac{4\sqrt{3}}{27}\pi
\left (\lambda -\frac{3}{2^{{2}/{3}}}\right ),
\label{okolo0}
\end{equation}
and
the derivative $\partial_{E}\tau |_{E_{max}}$ is indeed negative
\begin{equation}
{\partial\tau}/{\partial E}\bigg |_{E_{max}} = -\frac{8\sqrt{3}\pi}{27}
\frac{\alpha B^{{1}/{3}}}{F^{{4}/{3}}}.
\end{equation}
After solving the equation $\tau (E)={\hbar}/{T}$
in the vicinity of the point $E=E_{\rm max}$ 
we obtain the expression for the action
\begin{equation}
S_{Eucl}(T)\simeq\frac{\hbar}{T}
\left (U_{0}-\frac{3}{2^{{2}/{3}}}B^{{1}/{3}}F^{{2}/{3}}\right )-
2^{{4}/{3}}\sqrt{3}\pi^{3}\frac{\alpha^{3}B}{\hbar^{2}F^{4}}
{\left (T-T_{c}\right )}^{2}\theta\left (T_{c}-T\right )
\label{vicinity}
\end{equation}
which improves on the result (\ref{otwet}) in the vicinity
of $T_{c}.$

\vskip0.0cm
\centerline{\epsfxsize=16cm \epsfbox{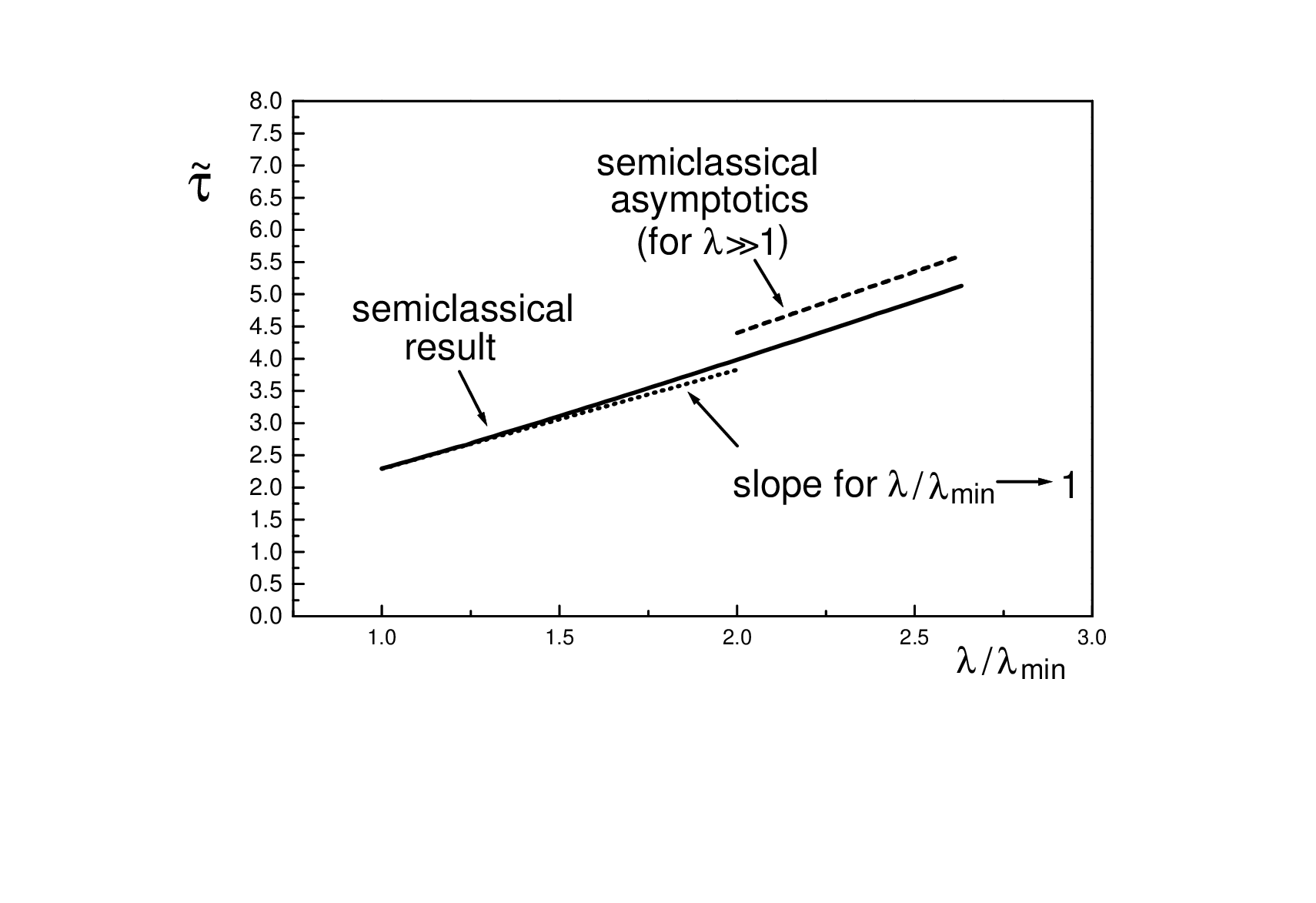}}
\vskip-2.5cm{\footnotesize {\bf Fig.4}~The function
${\tilde \tau}$ (solid line, see Eq.~(\ref{crossover1})),
which is directly proportional to the 
imaginary time oscillation period $\tau (E),$ 
as a function of
${\lambda}/{\lambda_{\rm min}}$ (see Eq.~(\ref{lambda})),
$\lambda_{\rm min}={3}/{2^{{2}/{3}}}.$ 
Surprisingly,
the slope of ${\tilde \tau}$ near 
${\lambda_{\rm min}}$ is very close to its asymptotic 
value at large $\lambda.$
The dotted line shows the function
$g(\lambda )= 
\frac{2^{{1}/{3}}}{\sqrt{3}}\pi+\frac{4\sqrt{3}}{27}\pi 
\left (\lambda -\frac{3}{2^{{2}/{3}}}\right ),$  
see Eq.~(\ref{okolo0}). 
The dashed line illustrates the slope of the function 
$f(\lambda )=\lambda + {Const }$ corresponding
 to the asymptotic expression
for $\tau (E),$
$\tau (E)\simeq{2\alpha\left (U_{0}-E\right )}/{F^{2}}.$
Obviously, the slope of $f(\lambda )$ is very close
to those of ${\tilde \tau } (\lambda )$ and $g(\lambda ),$
indicating that one can use Eq.~(\ref{period1})
 in almost the whole energy range.}

\subsection{Preexponential factor}

Finally, let us estimate the preexponential factor
at high and at low temperatures.
At $T>T_{c}$ the saddle-point solution is time independent
and is given by the equation $x=x_{max}$ with $x_{max}$
the maximum of the function $H(x,0).$
In the vicinity of the point $(x_{max},0),$
\begin{equation}
H\left (x, p\right )\simeq U_{0}-
\frac{3B}{x_{max}^{4}}
{\left (x-x_{max}\right )}^{2}+
\frac{B}{\alpha^{2}x_{max}^{4}}p^{2}.
\label{realtime}
\end{equation}
The imaginary time Hamiltonian corresponding to 
(\ref{realtime}) has the form of a harmonic oscillator
with the mass 
${{\alpha}^{2}x_{max}^{4}}/{2B}$ 
and the stiffness
${6B}/{x_{max}^{4}}.$ Consequently, at high temperatures
one can use the result for the decay rate of a massive particle
(see Refs.\cite{Haenggi,Affleck})
\begin{equation}
\Gamma=
\frac{\sqrt{3}F^{{4}/{3}}}{2^{{4}/{3}}\pi\alpha B^{{4}/{3}}}
\frac{\sinh {\left (\frac{\hbar\kappa}{2\alpha T}\right )}}
{\sin{ \left (
{\frac{\sqrt{3}}{2^{{4}/{3}}}\frac{\hbar F^{{4}/{3}}}{\alpha B^{{4}/{3}}T}}
\right )}}
\exp{\left (-\frac{U_{0}}{T}\right )}.
\end{equation}  
This expression is applicable at any temperature
higher than $T_{c}$ except for a narrow temperature interval 
$\sim {\hbar}^{{3}/{2}}$
around $T_{c},$ see Ref.\cite{Affleck}.
At $T\gg T_{c},{\hbar\kappa}/{\alpha}$ 
we obtain the simple result
$\Gamma =\left ({\hbar \kappa}/{2\pi\alpha}\right )
\exp{\left (-{U_{0}}/{T}\right )}.$

At zero temperature the decay involves only tunneling
out of the ground state, i.e.,
$\Gamma \simeq\left [{\omega (E_{0})}/{2\pi}\right ]
\exp{\left [-\left ({\alpha}/{\hbar}\right ) 
{\left ({U_{0}}/{F}\right )}^{2}\right ]},$
see Ref.\cite{Lifshitz}.
Near the metastable minimum the vortex oscillates with the frequency
$\omega ={\kappa}/{\alpha},$ and we arrive at the result
\begin{equation}
\Gamma \simeq
\frac{\kappa}{2\pi\alpha}
\exp{\left [-\frac{\alpha}{\hbar}{\left (\frac{U_{0}}{F} \right )}^{2}\right ]}.
\label{nul}
\end{equation}
However, one should point out that the preexponential factor
in Eq.~(\ref{nul}) is only an order of magnitude estimate
as the quasiclassical approximation is not
properly applicable to  the ground state.

\section{Applicability}
\label{conclusion}

Let us discuss the conditions for the applicability of  the above results.
First of all, 
we have to
account for a  nonzero vortex mass. In this
case, the system is described by the Hamiltonian
\begin{equation}
{\hat H}=
\frac{
{{\hat P}_{x}}^{2}
+{({\hat P}_{y}^{2}+\alpha x )}^{2}}{2 m}+
U_{0}\left (\sqrt{x^{2}+y^{2}}\right )-Fx,
\label{magnpol}
\end{equation}
which is equivalent to the Hamiltonian of a massive 2D charged particle in a
magnetic field 
oriented 
perpendicular to the plane of motion.
Let us discuss the conditions which should be satisfied in order to use
the Hamiltonian (\ref{Hamiltonian}) instead of that given by 
Eq.~(\ref{magnpol}). 
It is useful to introduce the four new operators (see also
Ref.\cite{Dmitriev}) 
${\hat \xi}_{y}=-{l^{2}{\hat P}_{x}}/{\hbar},$
${\hat X}=-{{l^{2}{\hat P}_{y}}}/{\hbar},$ 
${\hat \xi}_{x}={l^{2}{\hat P_{y}}}/{\hbar}+x,$ 
${\hat Y}={l^{2}{\hat P}_{x}}/{\hbar}+y,$
where $l=\sqrt{{\hbar}/{\alpha}}$ is the magnetic length\cite{Landau}.
We assume that the condition $l\ll a$ 
is satisfied (below we 
will
verify
 this condition using parameters for high-$T_{c}$-superconductors).
These new operators satisfy the commutation relations
$[{\hat\xi}_{y},{\hat \xi}_{x}]=[{\hat X},{\hat Y}]=il^{2},$
$[{\hat\xi}_{x},{\hat Y}]=[{\hat \xi}_{x},{\hat X}]=[{\hat\xi}_{y},{\hat X}]=
[{\hat\xi_{y}},{\hat Y}]=0.$ 
Going over to
the $X,\xi_{y}$-representation 
the Hamiltonian~(\ref{magnpol}) 
can be written in the form
\begin{equation}
{\hat H}=-\frac{\hbar^{2}}{2m}\frac{\partial^{2}}{\partial\xi_{y}^{2}}
+\frac{m\omega_{c}^{2}\xi_{y}^{2}}{2}+
\int U({\bf k})e^{ik_{x}({\hat \xi_{x}}+{\hat X})+
ik_{y}({\hat \xi_{y}}+{\hat Y})}d^{2}k.
\label{Hammmm}
\end{equation}
In a classical language
the problem we study involves two types of motion:
A fast rotation of the particle with the cyclotron frequency 
$\omega_{c}={\alpha}/{m}$
superimposed on 
the slow guiding center motion. The term
\begin{equation}
{\hat h }=-\frac{\hbar^{2}}{2m}\frac{\partial^{2}}{\partial\xi_{y}^{2}}+
\frac{m\omega_{c}^{2}\xi_{y}^{2}}{2}
\end{equation}
 describes the 
``fast'' part of the Hamiltonian. The lowest eigenvalue of ${\hat h}$
is equal to ${{\hbar}\omega_{c}}/{2}.$ If the characteristic
variation of the potential on the length $l$ is much smaller
than ${\hbar\omega_{c}}/{2} ,$ we can average
the Hamiltonian (\ref{Hammmm}) over the groundstate eigenfunction
$\chi (\xi_{y})$ of the operator ${\hat h},$
see Ref.\cite{note}.
 This process corresponds to the
averaging over the fast rotation with the frequency $\omega_{c}.$
After averaging we obtain
\begin{equation}
{\hat H}\rightarrow {\langle {\hat H}\rangle }_{\rm fast} =
\int U({\bf k})\exp{\left [-\frac{{\bf k}^{2}l^{2}}{4}
+i\left (k_{x}{\hat X}+k_{y}{\hat Y}\right)\right ]}d^{2}k.
\label{finalpotent}
\end{equation}
Note that
in the $X$-representation ${\hat Y}=-il^{2}{\partial}/{\partial X}.$
If $l\ll a,$ we can neglect the exponent $\exp{(-{{\bf k}^{2}l^{2}}/{4})}$
in Eq.~(\ref{finalpotent}).
On the other hand, if $l$ is small,
$[{\hat X},{\hat Y}]\rightarrow 0$ and we
can easily perform the integration in Eq.~(\ref{finalpotent})
to arrive at the new Hamiltonian
\begin{equation}
{\hat H}_{\rm eff}
=U_{0}\left (\sqrt{{\hat X}^{2}+{\hat Y}^{2}}\right )-F{\hat X}
\label{konez}
\end{equation}
describing the guiding
center motion. As the operators ${\hat X}$ and ${\hat Y}$ obey 
the commutation
relation $[{\hat X},{\hat Y}]=i l^{2},$ we can identify
${\hat X}$ with the coordinate and ${\hat Y}$ with the momentum. 
This Hamiltonian is identical to that in Eq.~(\ref{Hamiltonian}).

In order to carry out the above procedure
the characteristic variation of the potential $U(x,y)$ on the scale
$l$ 
has to be
much smaller than ${\hbar\omega_{c}},$ i.~e., 
$\kappa l^{2}\ll {\hbar\omega_{c}},$ and with $\kappa\sim {U_{0}}/{a^{2}}$
we obtain the condition ${\alpha^{2}a^{2}}/{m U_{0}}\gg 1.$  
Consequently, we arrive at the following two conditions for the 
applicability of the approach we have used,
\begin{equation}
l\ll a\ \ \ \ \ {\rm and}\ \ \ \ \  \frac{\alpha^{2}a^{2}}{mU_{0}}\gg 1.
\label{criterionnn}
\end{equation}
Let us
estimate the parameters in Eq.~(\ref{criterionnn})
appropriate for high-$T_{c}$-superconductors:
$\alpha =\pi\hbar  ns,$ where 
$n={ 2\cdot 10^{21} {\rm cm}^{-3}}$
is the electron density and $s=15\ {\rm \AA }$ is the 
interlayer spacing in Bi-2:2:1:2. For the magnetic length we then obtain
$l=\sqrt{{\hbar}/{\alpha}}\simeq{\rm 4\ \AA}\ll a\simeq {\rm 30\ \AA}.$
For the second condition in Eq.~(\ref{criterionnn}) we use
the expression for
the  vortex mass predicted for the superclean limit\cite{Simanek} 
$m=m_{e}{\left ({\epsilon_{F}}/{\Delta}\right )}^{2}\simeq 100m_{e}$ 
and the depth of the 
potential as given by the equation
$U_{0}={\left ({\Phi_{0}}/{4\pi\lambda_{ab}}\right )}^{2}s ,$
with $\Phi_{0}$ the magnetic flux quantum and 
$\lambda_{ab}=2000\ {\rm \AA}$ the $ab$-magnetic penetration length.
Finally, we obtain ${\alpha^{2}a^{2}}/{mU_{0}}\simeq 10\gg 1,$
i.e., we find that 
both conditions 
in Eq.~(\ref{criterionnn})
are well satisfied. 

The above results should be compared to those 
 in Ref.\cite{Bulaevskii} where the
condition ${\hbar\alpha}/{m U_{0}}\gg 1$ 
for the applicability 
of the results has been used. Estimating this parameter
we obtain ${\hbar\alpha}/{m U_{0}}\simeq 0.1\ll 1.$ In Ref.\cite{Bulaevskii}
an estimate  
 ${\hbar\alpha}/{m U_{0}}\simeq 200$ has been obtained
using the vortex mass from the dirty limit\cite{Suhl,Kupriyanov,Geshkenbein}, 
which is much smaller than the mass in the superclean limit.
We see that the LLL-approach of Ref.\cite{Bulaevskii}
breaks down in the superclean limit if one does not take 
explicitly the condition
$l\ll a$ into account.

Finally, let us show that the second condition in Eq.~(\ref{criterionnn})
can be easily obtained from a simple analysis of  the semiclassical 
bounce trajectories:
If the Euclidean action is much larger than unity, we can use
the instanton method for the calculation of the imaginary part
of the partition function which determines
the decay rate $\Gamma\propto {{\rm Im}Z}/{Z}.$
In order to determine ${\rm Im}Z$ within exponential accuracy
we have to solve the classical equation of motion describing
the bounce solution. If it turns out that the correction
of the bounce trajectory
due to the mass term is small, we arrive at an 
effective
Hall tunneling problem.
It is possible to neglect the mass if 
everywhere along the trajectory
$m\left |{\dot v}\right |\ll \alpha |v|,$
where ${\dot v}$ is the characteristic acceleration of the vortex
during the imaginary time motion and $v$ its characteristic velocity.
If 
a vortex is moving along the sides $CA, AA^{\prime},$ or $A^{\prime}C$
of the triangle $CAA^{\prime}$ (see Fig.~2), 
we can make  the  estimate 
$|{\dot v}|\sim v^{2}/{\left ({U_{0}}/{F}\right )},$ whereas for
$v$ the estimate  $v\sim {U_{0}}/{\alpha a}$ holds,
i.e., the following condition should be satisfied
\begin{equation}
\frac{F}{F_{c}}\ll \frac{{\alpha}^{2}a^{2}}{m U_{0}},
\label{classics}
\end{equation}    
with $F_{c}\sim {U_{0}}/{a}$  the critical force.
On the other hand, in the vicinity of the point
$C,$ we can write $v\sim{U_{0}}/{\alpha a}$ and
${\dot u}\sim v^{2}/{a},$ i.e., the condition 
${\alpha^{2}a^{2}}/{m U_{0}}\gg 1$ 
has to be satisfied,
which is identical
to the second condition in Eq.~(\ref{criterionnn}).
 Comparing this result with Eq.~(\ref{classics})
we see that the second condition in Eq.~(\ref{criterionnn})
is stronger since $F<F_{c}.$

\section{Summary} 
 In section \ref{conclusion} we have described how to reduce
the dynamics of a massive particle in a magnetic field 
to the guiding center motion along equipotential
lines of a smooth potential. Carrying out this procedure for vortices is not
entirely unproblematic. In fact, we have seen, that the averaging
process over the fast component introduces a frequency
$\omega_{c}={\alpha}/{m}\simeq\omega_{0}$
which is at the limit of the applicability of the low frequency vortex 
dynamics as described by Eq.~(\ref{dwizhenie}).
This problem has been ignored in previous works as the inertial
term has been dropped on the classical level of Eq.~(\ref{dwizhenie}).
As shown above, both methods produce the same condition
${{\alpha^{2}}a^{2}}/{m U_{0}}\gg 1$ for the irrelevance
of the mass term, however, starting from the description
(\ref{magnpol}) one obtains that the frequency $\omega_{c}={\alpha}/{m}$
naturally shows up in the quantum description.

Briefly summarizing, we have considered the decay
of the metastable state of a massless particle in a magnetic field
trapped in a cylindrical attractive potential 
and subject to a small external force. The above problem is an appropriate model
for the depinning of pancake vortices in superclean Josephson-coupled
 superconductors
for the case of small ($j\ll j_{c}$) transport currents.
The Euclidean action as a function of
temperature is given by Eq.~(\ref{otwet});
For $T<T_{1}$ the action is constant up
to exponentially small corrections and decreases smoothly
above $T_{1}.$ At $T=T_{c},$
the crossover to the calssical regime takes place.
Close to and above $T_{c}$ 
the decay rate can be accurately described by  
Eq.~(\ref{vicinity}) which
reduces to the exact classical result above $T_{c}.$
For the potential satisfying the conditions
{\it i}) $U(r)$ is a monotonous function of $r$ and 
{\it ii}) $U(r)\cong U_{0}-{B}/{r^{2}},\thinspace r\rightarrow\infty ,$ 
the transition from quantum to classical
behavior is second-order like and the crossover temperature $T_{c}$ 
is proportional to $F^{{4}/{3}},$ see Eq.~(\ref{Tc}).

\acknowledgments
We thank L.~N.~Bulaevskii, V.~B.~Geshkenbein, B.~I.~Ivlev, N.~B.~Kopnin,
and V.~M.~Vinokur
 for helpful discussions.


\begin{thebibliography}{99}
\bibitem{experim}
L.~Civale, A.~D.~Marwick, T.~K.~Worthington, M.~A.~Kirk,
J.~R.~Thompson, L.~Krusin-Elbaum, Y.~Sun, J.~R.~Clem,
and F.~Holtzberg, Phys. Rev. Lett. {\bf 67}, 648 (1991);
V.~Hardy, D.~Groult, J.~Provost, M.~Hervieu, and R.~Raveau, Physica
(Amsterdam) C {\bf 178}, 225 (1991); W.~Gerh\"auser, G.~Ries,
H.~W.~Neum\"uller, W.~Schmidt, O.~Elibi, G.~Saemann-Ischenko,
and Klaum\"unzer, Phys. Rev Lett. {\bf 68}, 879 (1992);
D.~Prost, L.~Fruchter, I.~A.~Campbell, N.~Motohira, and M.~Konczykowski,
Phys. Rev. B {\bf 47}, 3457 (1993).
\bibitem{Bulaevskii}
L.~N.~Bulaevskii, A.~I.~Larkin, M.~P.~Maley, and
V.~M.~Vinokur, Phys. Rev. B {\bf 52}, 9205 (1995).
\bibitem{Haenggi}P.~H\"anggi, P.~Talkner, and B.~Borkovec, Rev. Mod.
Phys. {\bf 62}, 251 (1990).
\bibitem{Feigel'man}M.~V.~Feigel'man, V.~B.~Geshkenbein,
A.~I.~Larkin, and S.~Levit, Pis'ma Zh. Eksp. Teor. Fiz. {\bf 57}, 699
(1993) [JETP Lett. {\bf 57}, 711 (1993)].
\bibitem{Ao}P.~Ao and D.~J.~Thouless, Phys. Rev. Lett. {\bf 72},
132 (1994).
\bibitem{Stephen}M.~J.~Stephen, Phys. Rev. Lett {\bf 72}, 1534
(1994).
\bibitem{Chudnovsky3}E.~M.~Chudnovsky, A.~Ferrera, and A.~Vilenkin,
Phys Rev. B {\bf 51}, 1181 (1995).
\bibitem{Sonin}E.~B.~Sonin and B.~Horovitz, Phys. Rev. B {\bf 51},
6526 (1995).
\bibitem{Morais-Smith}C.~Morais-Smith, A.~O.~Caldeira, and
G.~Blatter,
Phys. Rev. B {\bf 54}, 784 (1996).
\bibitem{Gorokhov2}D.~A.~Gorokhov and G.~Blatter, 
``Thermally activated Hall creep of flux lines from
a columnar defect'', accepted for publication in Phys. Rev. B.
\bibitem{Blatter}G.~Blatter, M.~V.~Feigel'man,
V.~B.~Geshkenbein, A.~I.~Larkin, and V.~M.~Vinokur,
Rev. Mod. Phys. {\bf 66}, 1125 (1994). 
\bibitem{Matsuda}Y.~Matsuda, N.~P.~Ong, Y.~F.~Yan, J.~M.~Harris,
and J.~B.~Peterson, Phys. Rev. B {\bf 49}, 4380 (1994).
\bibitem{Harris}J.~M.~Harris, Y.~F.~Yan, O.~K.~C.~Tsui,
Y.~Matsuda, and N.~P.~Ong, Phys. Rev. Lett. {\bf 73}, 1711 (1994).
\bibitem{Kopnin}
N.~B.~Kopnin and V.~E.~Kravtsov, 
Pis'ma Zh. Eksp. Teor. Fiz. {\bf 23}, 631 (1976) [JETP Lett.
{\bf 23}, 578 (1976)]; Zh. Eksp. Teor. Fiz. {\bf 71}, 1644 (1976)
[Sov. Phys. JETP {\bf 44}, 861 (1976)];
N.~B.~Kopnin and M.~M.~Salomaa, Phys. Rev. B {\bf 44},
9667 (1991).
\bibitem{Simanek}E.~\u{S}im\'anek, Phys. Lett. A {\bf 194}, 323
(1994).
\bibitem{AVO}A.~van~Otterlo, M.~Feigel'man, V.~Geshkenbein, and G.~Blatter,
Phys. Rev. Lett. {\bf 75}, 3736 (1995).
\bibitem{Mkrtchyan}G.~S.~Mkrtchyan and V.~V.~Shmidt, Zh. Eksp. Teor. Fiz.
{\bf 61}, 367 [Sov Phys. JETP {\bf 34}, 195 (1972)].
\bibitem{note1}In the present problem 
the saddle-point solution is always located in the sector $|x|>y.$ 
\bibitem{Volovik}G.~E.~Volovik, Pis'ma Zh. Eksp. Teor. Fiz. {\bf 15},
116 (1972) [JETP Lett. {\bf 15}, 81 (1972)].
\bibitem{Landau}L.~D.~Landau and E.~M.~Lifshitz,
{\it Quantum Mechanics}, Course in Theoretical Physics, Vol.~3
(Pergamon, Oxford, 1977).
\bibitem{Galitsky}V.~ M.~Galitsky, B.~M.~Karnakov, and V.~I.~Kogan,
{\it Zadachi po kvantovoi mechanike}, [in Russian] 
({\it Problems in Quantum Mechanics}) (Nauka, Moscow, 1992).
\bibitem{Grabert}H.~Grabert and U.~Weiss, Z. Phys. B {\bf 56}, 171 (1984).
\bibitem{Lifshitz}
I.~M.~Lifshitz and Yu.~Kagan, Zh. Eksp. Teor. Fiz. {\bf 62}, 1
(1972) [Sov. Phys. --- JETP {\bf 35}, 206 (1972)].
\bibitem{Chudnovsky}
S.~V.~Meshkov, Zh. Eksp. Teor. Fiz. {\bf 89}, 1734 (1985)
[Sov. Phys. JETP {\bf 62}, 1000 (1985)];
A.~S.~Ioselevich and E.~I.~Rashba, Zh. Eksp. Teor. Fiz. {\bf 91},
1917 (1986) [Sov. Phys. JETP {\bf 64}, 1137 (1986)].
\bibitem{Chudnovsky1}E. M. Chudnovsky, Phys. Rev. A {\bf 46}, 8011 (1992).
\bibitem{Gorokhov} D.~A.~Gorokhov and G.~Blatter,
Phys. Rev. B {\bf 56}, 3130 (1997). 
\bibitem{Affleck}
I.~Affleck, Phys. Rev. Lett. {\bf 46}, 388 (1981). 
\bibitem{Dmitriev}A.~P.~Dmitriev and V.~Yu.~Kachorovskii,
Phys. Rev. B {\bf 52}, 5743 (1995).
\bibitem{note}High Landau levels also contribute to the decay
rate, however, the Euclidean action involved is larger.
\bibitem{Suhl}H.~Suhl, Phys. Rev. Lett. {\bf 14}, 226 (1965).
\bibitem{Kupriyanov}M.~Yu.~Kupriyanov and K.~Likharev,
Zh. Eksp. Teor. Fiz. {\bf 68}, 1506 [Sov. Phys. JETP {\bf 41}, 755 (1975)].
\bibitem{Geshkenbein}G.~Blatter, V.~B.~Geshkenbein, and V.~M.~Vinokur,
Phys. Rev. Lett {\bf 66}, 3297 (1991).





\end{thebibliography}
\end{document}